\documentclass[11pt]{article}
\usepackage{ol2}
\usepackage[draft]{hyperref}
\usepackage{amsmath}
\usepackage{graphicx}          
\usepackage{dcolumn}
\usepackage{amssymb}
\usepackage{color}
\usepackage{epstopdf}

\newcommand{\ba}{\begin{array}}
\newcommand{\ea}{\end{array}}
\newcommand{\be}{\begin{equation}}
\newcommand{\ee}{\end{equation}}
\newcommand{\bea}{\begin{eqnarray}}
\newcommand{\eea}{\end{eqnarray}}

\newcommand{\bfj}{{\bf j}}

\newcommand{\bfx}{{\bf x}}

\newcommand{\bfz}{{\bf z}}

\newcommand{\bfp}{{\bf p}}

\newcommand{\bfF}{{\bf F}}
\newcommand{\bfG}{{\bf \Gamma}}

\begin{document}


\title{Macroscopic direct observation of optical spin-dependent\\lateral forces and left-handed torques}

\vspace{-8mm}
\author{Hernando Magallanes and Etienne Brasselet*} \vspace{1mm}
\address{Univ. Bordeaux, CNRS, Laboratoire Ondes et Mati{\`e}re d'Aquitaine,\\ 351 cours de la lib\'eration, F-33400 Talence, France.\\
*Corresponding author: etienne.brasselet@u-bordeaux.fr}
\vspace{4mm}

{\bf
Observing and taming unusual effects arising from non-trivial light-matter interaction has always triggered scientists to better understand Nature and develop technological tools towards implementing novel applications. Recently, several unusual optomechanical effects have been unveiled when subtle spin-orbit interactions come at play to build up optical forces and torques that are hardly seen in everyday life, such as negative optical radiation pressure, transverse optical forces, or left-handed optical torques. To date, there are only a few experimental proposals to reveal these effects \cite{brzobohaty_np_13, dogariu_np_13, sukhov_natphot_2015, antognozzi_natphys_2016, hakobyan_natphot_2014} despite tremendous conceptual advances \cite{bliokh_natphot_2015, sukhov_rpp_2017}. In particular, spin-dependent lateral forces and their angular analog are done either at the expense of direct observations or at the cost of specific instrumental complexity. Here we report on naked-eye identification of light-induced spin-dependent lateral displacements of centimeter-sized objects endowed with structured birefringence. Its angular counterpart is also discussed and the observation of left-handed macroscopic rotational motion is reported. The unveiled effects are ultimately driven by lateral optical force fields that are five orders of magnitude larger than those reported so far. These results allow structured light-matter interaction to move from a scientific curiosity to a new asset for the existing multidisciplinary optical manipulation toolbox across length scales \cite{grier_review_03, jonas_electrophoresis_2008, bekshaev_jo_2013, padgett_review_twist11, gao_lsa_2017}. In addition, this highlights the spin-orbit optomechanics of anisotropic and inhomogeneous media, which is just beginning to be explored \cite{cipparrone_pra_2011, angelsky_oe_2017, tkachenko_pra_2017}.
}

In free space, paraxial light carries longitudinal linear and angular momenta, both pointing along its propagation direction as their name suggests. The collinearity between the incident momenta and the direction of the radiation forces and torques exerted by light on matter is not a given, as emphasized by the case study of a beam reflecting off a perfect mirror illustrated in Fig.~1. An optical force $\bfF$ appears in a direction perpendicular to the surface of the mirror and always points towards it, see Fig.~1(a). On the other hand, an optical torque $\bfG$ appears in a direction parallel of the surface of the mirror with an orientation that depends on the projection of the incident angular momentum along the incident propagation direction, see Fig.~1(b). It is well understood that both $\bfF$ and  $\bfG$ lie in the incidence plane $(x,z)$ and their evaluation follows the simple use of Newton's laws of mechanics. However, the situation becomes much more complex when an object is placed nearby an interface, even in the simplest case of planar surface separating two semi-infinite media, for which transverse optical forces and torques with respect to the driving flow of light can appear. For instance, this has been discussed theoretically for particles placed in the evanescent field of dielectric \cite{bliokh_natcomm_2014} or plasmonic \cite{canaguier_pra_2014} interfaces, soon followed by the first experimental demonstrations of spin-dependent lateral forces \cite{sukhov_natphot_2015, antognozzi_natphys_2016}. Besides the case of  pure surface waves, the back-action of guided modes in presence of material spatial confinement has also been explored for circularly oscillating light-induced \cite{rodriguez_natcomm_2015} or spontaneously-emitting\cite{scheel_pra_2015} dipoles. We note that the presence of a tangible light source is not a prerequisite to lateral electromagnetic forces, as recently demonstrated in the case of a rotating particle near a planar surface experiencing lateral Casimir effects \cite{manjavacas_prl_2017}. Other situations where lateral forces appear have been proposed using purely optical schemes, by placing a spinning particle in the course of a light beam  \cite{chang_jjap_1994, movassagh_pra_2013}, or placing a particle in a inhomogeneous light field resulting two-wave interferences \cite{sukhov_optica_2014, bekshaev_prx_2015, fardad_ol_2016} or tightly focused vector beams \cite{li_pra_2017}. Finally, we also mention many theoretical efforts to deploy spin-dependent lateral forces towards sorting material chirality by chiral light \cite{shang_oe_2013, wang_natcomm_2014, cameron_njp_2014, hayat_pnas_2015, canaguier_jo_2015, alizadeh_ACSphot_2015, chen_pra_2016, zhang_acsnano_2017, cao_nanoscale_2018} soon after first experimental demonstrations restricted to the use of longitudinal forces \cite{tkachenko_prl_2013, tkachenko_natcomm_2014_sorting}. That said, the study of lateral optical forces represents an attractive emerging research area with both fundamental and applied interests.

\begin{figure}[t!]
\centering\includegraphics[width=0.5\columnwidth]{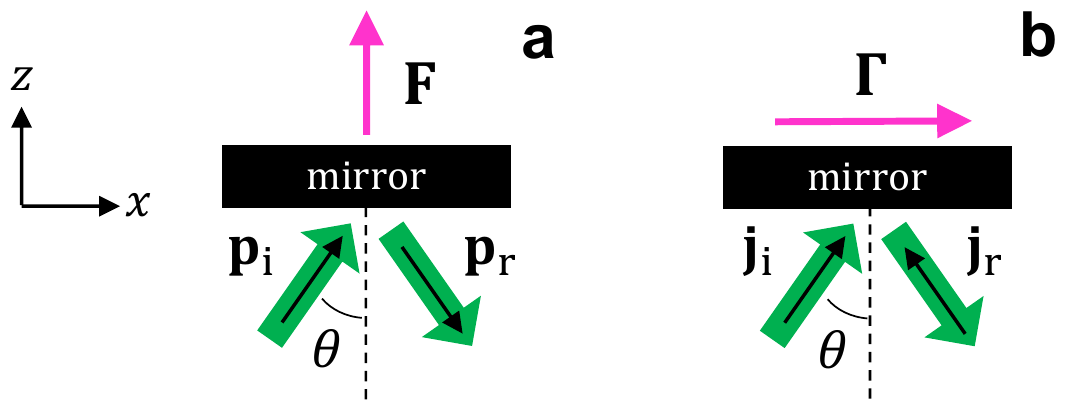}
\caption{
{\small{\bf In-plane translational and rotational optomechanics from the reflection on a perfect mirror.} The balance of linear and angular momenta per photon, $\bfp$ and $\bfj$, between the incident (subscript 'i') and reflected (subscript 'r') fields as a paraxial light beam (green arrow) with power $P$ is reflected on a perfect mirror at an angle $\theta$ gives $\bfF = 2(P/c)\cos\theta\, \bfz$ (${\bf a}$) and  $\bfG = 2(jP/\omega)\sin\theta \,\bfx$ (${\bf b}$) with $c$ the speed of light, $j$ the magnitude of the angular momentum per photon in $\hbar$ units and $\omega$ the angular frequency of light. These examples illustrate that the mechanical action is not necessarily pointing along the momenta at the origin of the effect, though being mutually in-plane.}
}
\end{figure}

Surprisingly, there are very few optical experiments reporting on the direct observation of spin-dependent lateral  forces. This was achieved using either isotropic dielectric microparticles \cite{sukhov_natphot_2015}, planar dielectric nanocantilever \cite{antognozzi_natphys_2016}, or liquid crystal bipolar droplet \cite{cipparrone_pra_2011}---though the understanding of this last remains unclear to date. These spin-orbit optomechanical experiments are all based on micron-sized objects and involve force magnitude up to $1-10$~fN associated with lateral displacement up to $0.1-10~\mu$m. In particular, the weakness of the effects comes with instrumental difficulties. Namely, force enhancement via many-particle hydrodynamic interaction is needed in \cite{sukhov_natphot_2015}. On the other hand, polarization-resolved reconstruction of both the longitudinal and transverse force components based on symmetry considerations is needed in \cite{antognozzi_natphys_2016}. In contrast, here we report on the observation per se of the action of spin-dependent lateral force on a centimeter-sized transparent, anisotropic and inhomogeneous dielectric slab irradiated by a circularly polarized continuous-wave Gaussian laser beam at wavelength $\lambda = 532$~nm. The applied force is of the order of 1~nN and leads to lateral displacements over a distance up to 0.5~mm in less than 100~s for an incident optical power of 1~W. Watt for Watt, our macroscopic experiment thus deals with an increase of the force by five orders of magnitude with respect to previous works.


\begin{figure}[b!]
\centering\includegraphics[width=\columnwidth]{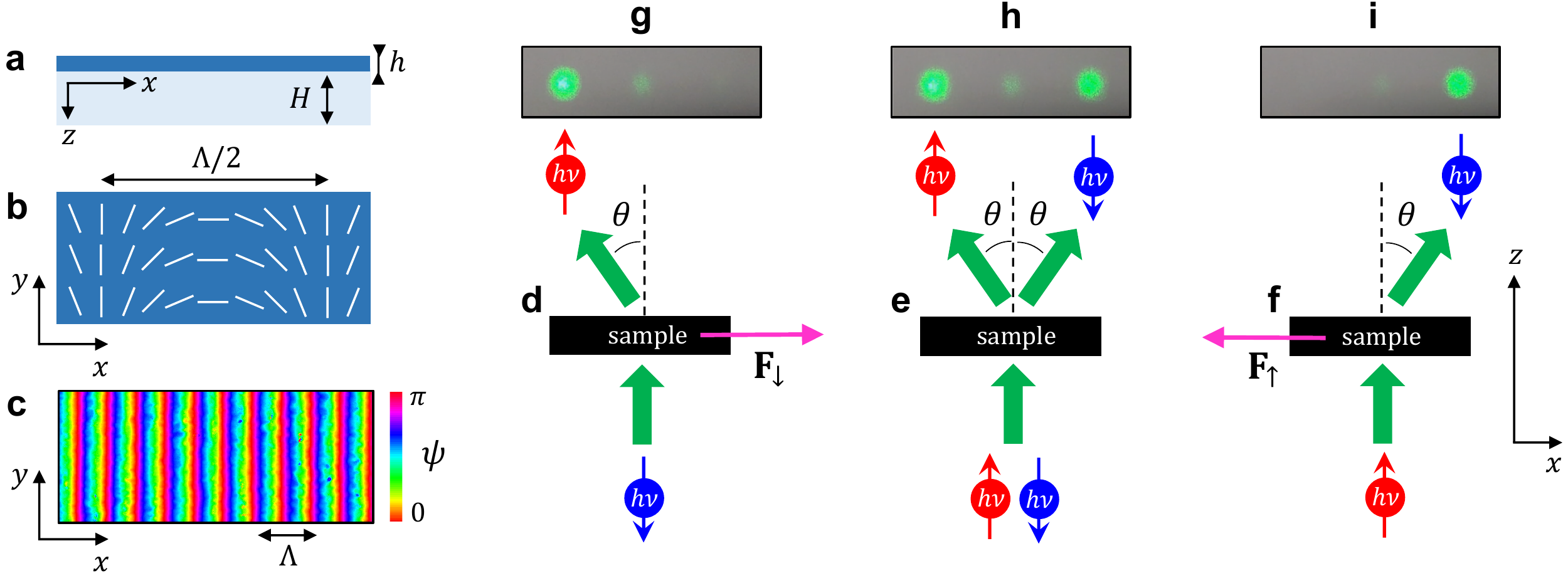}
\caption{
{\small{\bf Spin-dependent lateral optical forces from spin-orbit scattering by a one-dimensional geometric phase optical element.}
The material consists of a cycloidal diffractive waveplate with thickness $h\sim 2~\mu$m deposited on a transparent flexible plastic sheet of thickness $H \sim 200~\mu$m (${\bf a}$). Optical microscopy birefringent imaging reveals the one-dimensional periodic modulation of the optical axis orientation angle in the plane of the sample, here $\psi=-2\pi x/\Lambda$ (${\bf b}$,${\bf c}$). Spin-dependent beam steering associated with the deflection angle $\theta$ and spin-flipped transmitted photons leads to spin-dependent lateral optical forces $\bfF_\perp^{(\uparrow, \downarrow)}$ with opposite direction and same magnitude for incident spin up and down (${\bf d}$,${\bf f}$) and no net lateral force for incident linearly polarized light carrying equal weight fractions of spin up and down photons (${\bf e}$). The corresponding diffraction patterns are shown in (${\bf g}$)$-$(${\bf i}$).}
}
\end{figure}

Here, giant spin-dependent lateral optical force is achieved by coupling the incident spin angular momentum to the spatial degrees of freedom. Indeed, if `spin up' and `spin down' photons are selectively steered on either side of the direction of propagation of the incident light, spin-controlled reversal of the transverse linear momentum balance becomes possible, as illustrated in Fig.~2. Namely, we use cycloidal diffractive waveplates that are a transparent anisotropic (uniaxial) planar optical elements having uniform birefringent phase retardation $\Delta$ and one-dimensional space-varying optical axis orientation angle $\psi(x) = \pm2\pi x/\Lambda$, where $\Lambda$ refers to the distance over which the optical axis makes a full turn and $(x,y)$ defines the plane of the sample. These optical elements were introduced several decades ago and are characterized by the suppression of all the diffraction orders except the $\pm1$st ones provided that $\Delta=\pi$ \cite{nikolova_optacta_1984}. Later, they were recognized as geometric phase spin-dependent deflectors relying on the spin-orbit interaction of light \cite{bomzon_ol_2002}. Incident photons with spin up ($\uparrow$) and spin down ($\downarrow$) experience opposite geometric phase gradient associated with the deflection angle $\theta = \arcsin(2\lambda/\Lambda)$ and spin-flipped transmission. Here, the system is made of a structured liquid crystal polymer layer \cite{tabyrian_opn_2010} whose thickness and birefringence are designed to fulfill the half-wave birefringent retardation condition at $\lambda = 532~$nm, with $\Lambda \simeq 4.6~\mu$m, which gives $\theta \simeq 13^\circ$ in air. The structural characterization of the material is shown in Fig.~2(c) and its spin-dependent diffraction behavior is reported in Fig.~2(d-f) for an incident spin up and spin down states, and even-weight mixture of these. Ensuing direct observation of spin-dependent lateral forces $\bfF_\perp^{(\uparrow,\downarrow)}$, which satisfy $\bfF_\perp^{(\uparrow)} \cdot \bfF_\perp^{(\downarrow)} < 0$ and $|\bfF_\perp^{(\uparrow)}| = |\bfF_\perp^{(\downarrow)}|$, thus depends on the ability to design an experiment where the light-induced motion of the sample is not hindered.
 
\begin{figure}[b!]
\centering\includegraphics[width=1\columnwidth]{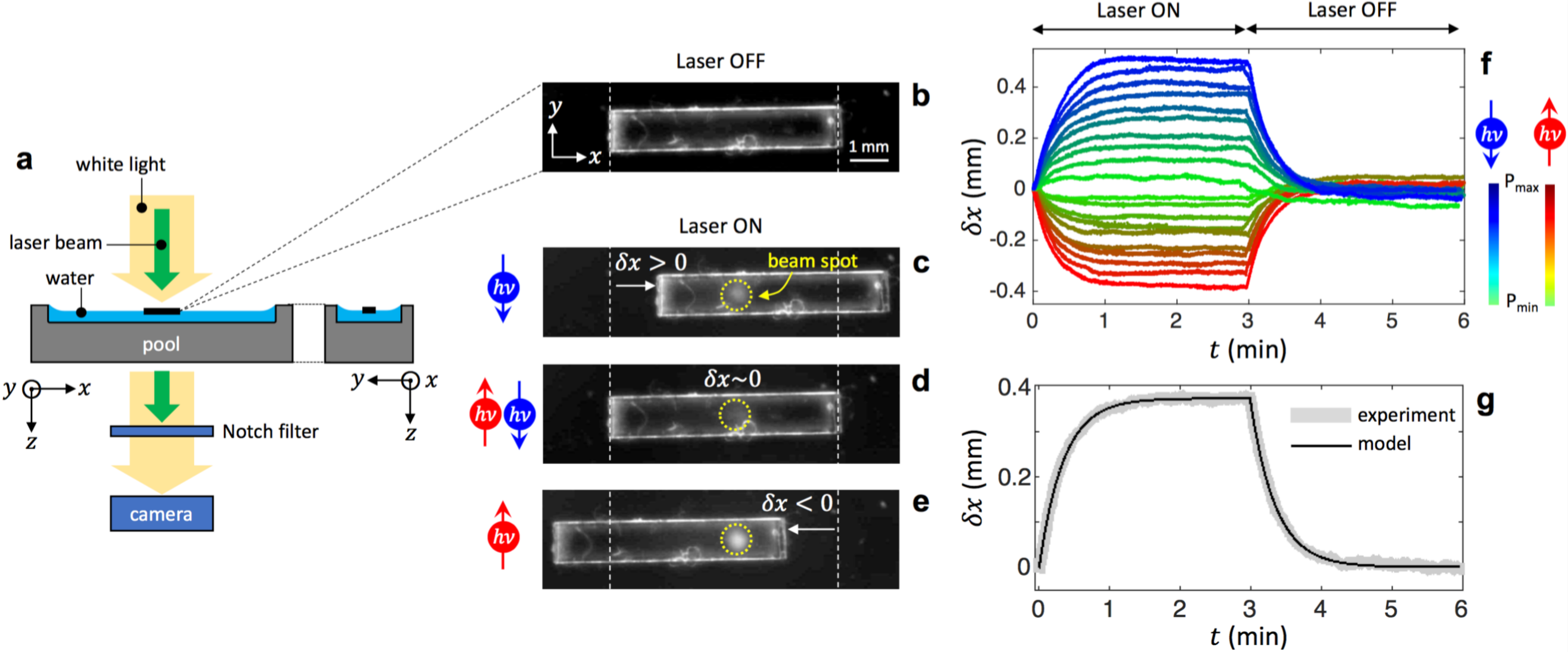}
\caption{
{\small {\bf Macroscopic direct observation of spin-dependent lateral optical forces.}
The sample is a $6~{\rm mm} \times 1~{\rm mm}$ rectangular slab hand-cut  from a large area cycloidal diffractive waveplate (see Fig.~2) placed at the air-water interface of a $7~{\rm cm} \times 1~{\rm cm}$ rectangular plexiglass pool with 5~mm depth (${\bf a}$). At rest, the sample is self-aligned and trapped in the center of the pool by capillary effects (${\bf b}$). Optomechanical effect is driven by a normally incident Gaussian laser beam with 0.9~mm waist diameter centered on the sample at rest. The sample dynamics is video recorded by a camera under unpolarized white light illumination with almost total extinction of the spin-orbit-scattered laser light by placing a spectrally selective so-called Notch filter for 532~nm wavelength in front of the camera. Typical steady-state spin-dependent displacement is shown in panels (${\bf c}$)$-$(${\bf e}$) under $\sim 1~$W incident optical power. (${\bf f}$) Power and spin dependence of excitation/relaxation dynamics for ten values of incident power between $P=0.14$~W and $P=1.4$~W by 0.14~W steps, for both spin up and spin down cases. Each curve is the average of three independent events. (${\bf g}$) Typical quantitative description if the translational dynamics by using a linear damped oscillator model, see Eqs.~(1)$-$(3).}
}
\end{figure}
 
As illustrated in Fig.~3(a), the sample is placed at air-water interface in a $7~{\rm cm} \times 1~{\rm cm}$ rectangular plexiglass pool with $5~{\rm mm}$ depth, by taking care to fix the solid-liquid-gas contact line at the rim of the slightly under-filled pool. The two main menisci with different radius of curvature thus ensure capillary alignment and stable trapping of a $6~{\rm mm} \times 1~{\rm mm}$ rectangular sample in the center of the pool as long as the laser beam is off as shown in Fig.~3(b) (see video 1 for launch and self-alignment of the sample). Qualitative optomechanical observations are reported in Figs.~3(c), 3(d) and 3(e) in presence of the laser beam with incident power $P\sim 1$~W for left-handed circular, linear, and right-handed circular incident polarization states, respectively (see videos 2-5). While the sample is not moving in the case of linear polarization, it moves and eventually stops as it gets closer to the meniscus for circular polarizations. Indeed, the meniscus behaves as a capillary spring loading with the sample displacement. Moreover, opposite lateral displacements with the same order of magnitude are observed when photons are prepared in either up or down pure spin states.

Quantitative analysis is made by recording the excitation and relaxation dynamics of the sample, $x(t)$. This is summarized in Fig.~3(f) for ten values of incident power between $P=0.14$~W and $P=1.4$~W, for both spin up and spin down cases. Each curve is the average of three on/off cycles for the laser irradiation, each cycle consisting of 3~min duration irradiation followed by 3~min duration relaxation. This protocol allows reliable data analysis as discussed hereafter. In fact, the simple yet relevant model of a linear forced oscillator with damping model is proposed, $m\ddot{x} + \gamma\dot{x} + \kappa x = H(t)F_\perp$,
where $m = 1.6$~mg is the mass of the sample, $\gamma$ refers to the losses arising mainly from the viscosity of water, $\kappa$ is the effective spring constant associated with the curved air-water interface, $H$ is a step function defined as $H(0<t<t_{\rm off}) = 1$ and $H(t>t_{\rm off}) = 0$, with the time zero taken as the moment when the laser is switched-on from the situation at rest, and $(\dot{x},\ddot{x})$ stands for $(dx/dt, d^2x/dt^2)$. According to the boundary conditions for the excitation and relaxation regimes (which assumes large enough $t_{\rm off}$ with respect to the transient excitation time), $\{x(0)=0,\dot{x}(0)=0\}$ and $\{x(t_{\rm off})=F_\perp/\kappa,\dot{x}(t_{\rm off})=0\}$, the equation of motion are $x_{\rm on}(0<t<t_{\rm off}) =  \frac{F_\perp}{\kappa}\,\left[1+\frac{\alpha_- e^{\alpha_{+} t} - \alpha_+ e^{\alpha_{-} t}}{\alpha_+ - \alpha_-}\right]$ and $x_{\rm off}(t>t_{\rm off}) = \frac{F_\perp}{\kappa}\,\frac{\alpha_+ e^{\alpha_{+} (t-t_{\rm off})} - \alpha_- e^{\alpha_{-} (t-t_{\rm off})}}{\alpha_- - \alpha_+}$,
%
with $\alpha_{\pm} = (-\gamma \pm\sqrt{\gamma^{2} - 4m\kappa})/(2m)$ and where we use the Minkowski expression for the spin-dependent lateral force, $F_\perp = \pm (P/c)\sin\theta$. A typical fit of the experimental data is shown in Fig.~3(g), where $\{\gamma, \kappa\}$ are the two adjustable parameters. Since the excitation an relaxation dynamics are processed independently for spin up and down, the full dataset provide us with 20 independent estimates both for $\{\gamma^{(\uparrow)}, \kappa^{(\uparrow)} \}$ and $\{\gamma^{(\downarrow)}, \kappa^{(\downarrow)} \}$, see Table 1. The obtained values for the viscous coefficient and the capillary spring constant for spin up and spin down point out the residual asymmetry of our macroscopic hand-cut sample and machined pool, yet not altering our conclusions.

\begin{table}[b!]
\centering
\begin{tabular}{ccc}
\hline
  & $\gamma$ (mg\,s$^{-1}$) &  $\kappa$ (mg\,s$^{-2}$)\\
\hline
spin up $(\uparrow)$  & $44.6 \pm 11.0$       & $2.17 \pm 0.24$ \\
spin down $(\downarrow)$ & $36.7 \pm 5.6$  & $1.92 \pm 0.09$ \\
\hline
\vspace{1mm}
\end{tabular}
\caption{
{\bf Data processing summary.}
{\small Estimated values for the viscous loss coefficient and the capillary spring constant obtained by adjusting all the experimental data presented in Fig.~2(f) with Eqs.~(2) and (3) for incident photons prepared in spin up and spin down states.}
}
\end{table}

Towards generalization of the above results, we note an optical torque that is collinear to the propagation direction of the incident beam is associated with a spatial distribution of transverse elementary forces. Therefore, since $\bfF_\perp \times {\boldsymbol \nabla} \psi = 0$, arbitrary spin-driven optical torques can be formally designed from appropriate material structuring. In particular, here we address the direct observation of a unusual effect earlier referred to the existence of a `left-handed' optical torque \cite{hakobyan_natphot_2014} where the applied total torque ${\boldsymbol \Gamma}$ and incident spin optical angular momentum ${\bf S}$ are anti-parallel, ${\bf\Gamma} \cdot {\bf S}<0$, in contrast to a `right-handed' torque where ${\bf\Gamma} \cdot {\bf S}>0$, as illustrated in Fig.~4. Early numerical prediction of left-handed torques has been reported in \cite{simpson_josaa_2007} using a wavelength-sized prolate particles made of transparent isotropic media under tightly focused circularly polarized Gaussian beams. Since then, several alternative routes have been discussed theoretically \cite{haefner_prl_2009, chen_scirep_2014, nieto_ol_2015, canaguier_pra_2015}. However, to date there is no direct experimental observation of such an effect.

\begin{figure}[t!]
\centering\includegraphics[width=0.55
\columnwidth]{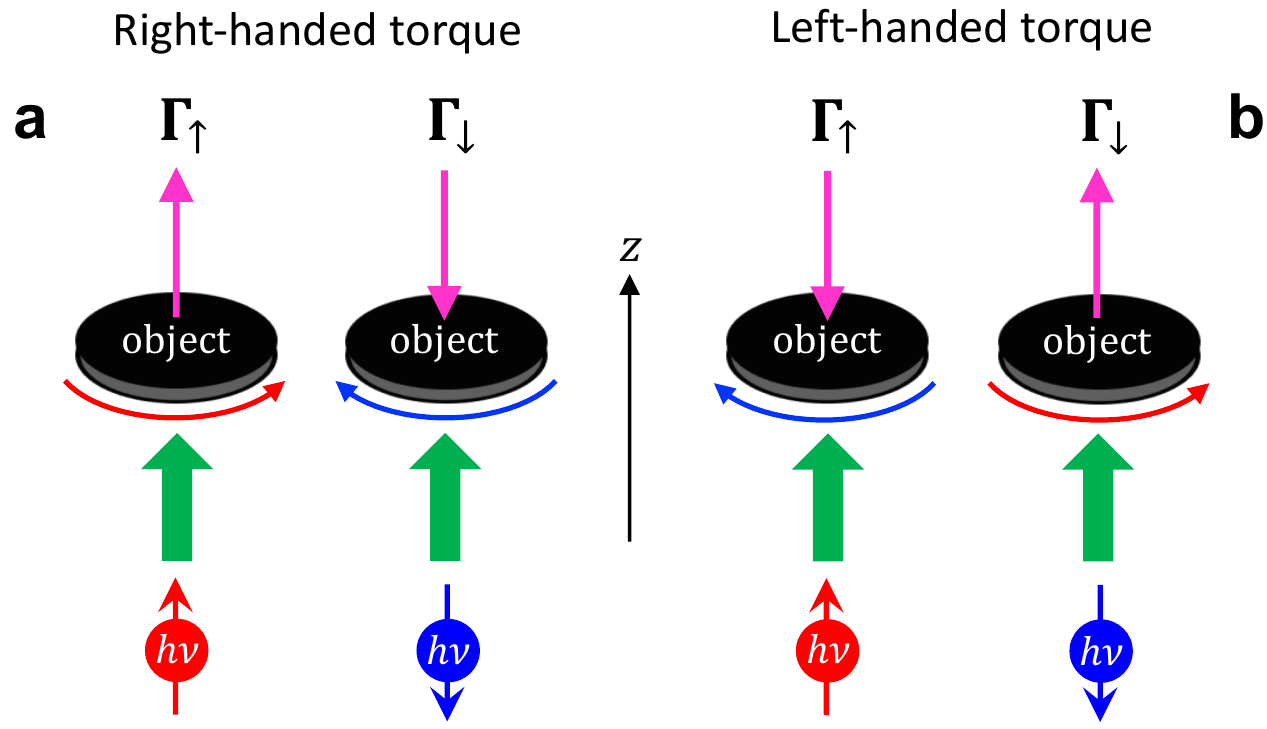}
\caption{
{\small {\bf Principle of right/left-handed optical radiation torques.}
An object subjected to the application of a usual optical torque (`right-handed') controlled by the angular momentum of spin tends to rotate in the same direction as that of the incident angular momentum (${\bf a}$). The reversed situation corresponds to unconventional (`left-handed') optical torque manifestation (${\bf b}$).}
}
\end{figure}

Here we propose to reverse the indirect approach reported in \cite{hakobyan_natphot_2014}, where the mechanical action of matter on light was analyzed, by directly looking at the mechanical action of light on matter. We fabricate $\sim 6~$mm diameter four-quadrant disk-shaped spin-orbit rotors from four hand-cut pieces of cycloidal diffractive waveplate according to the design shown in Fig.~5(a) and 5(b). The orientational gradients ${\boldsymbol \nabla} \psi$ are oriented counter-clockwise (right-handed case) or clockwise (left-handed case), respectively. In practice, the four pieces are deposited at the air-water interface of a $2~$cm diameter slightly under-filled 5~mm depth circular plexiglass pool where they remain bound via capillary effects. Recalling that ${\boldsymbol \Gamma} = \iint {\bf r} \times \bfF_\perp dxdy$ and $|\bfF_\perp| \propto I$ where $I$ refers to the spatial distribution of the light intensity, the torque magnitude is optimized by shaping the incident beam in a ring fashion using an axicon placed before a lens. Obtained axisymmetric intensity profile is described by three parameters ($\eta_0$, $r_0$, and $w_0$) according to the analytical expression $I(r)/I_{\rm max} = \eta_0 + (1-\eta_0)e^{-2(r-r_0)^2/w_0^2}$ for $r<r_0$ and $I(r)/I_{\rm max} = e^{-2(r-r_0)^2/w_0^2}$ for $r>r_0$, see Figs.~5(c)$-$5(e). This model allows to evaluate that our experimental conditions correspond to a torque magnitude $|\Gamma| =\frac{4\sin \theta}{c}\int_0^R I(r) r^2 dr$  that is $\sim 75\%$ of the expected ideal value, see Fig.~5(f),

\begin{figure}[h!]
\centering\includegraphics[width=\columnwidth]{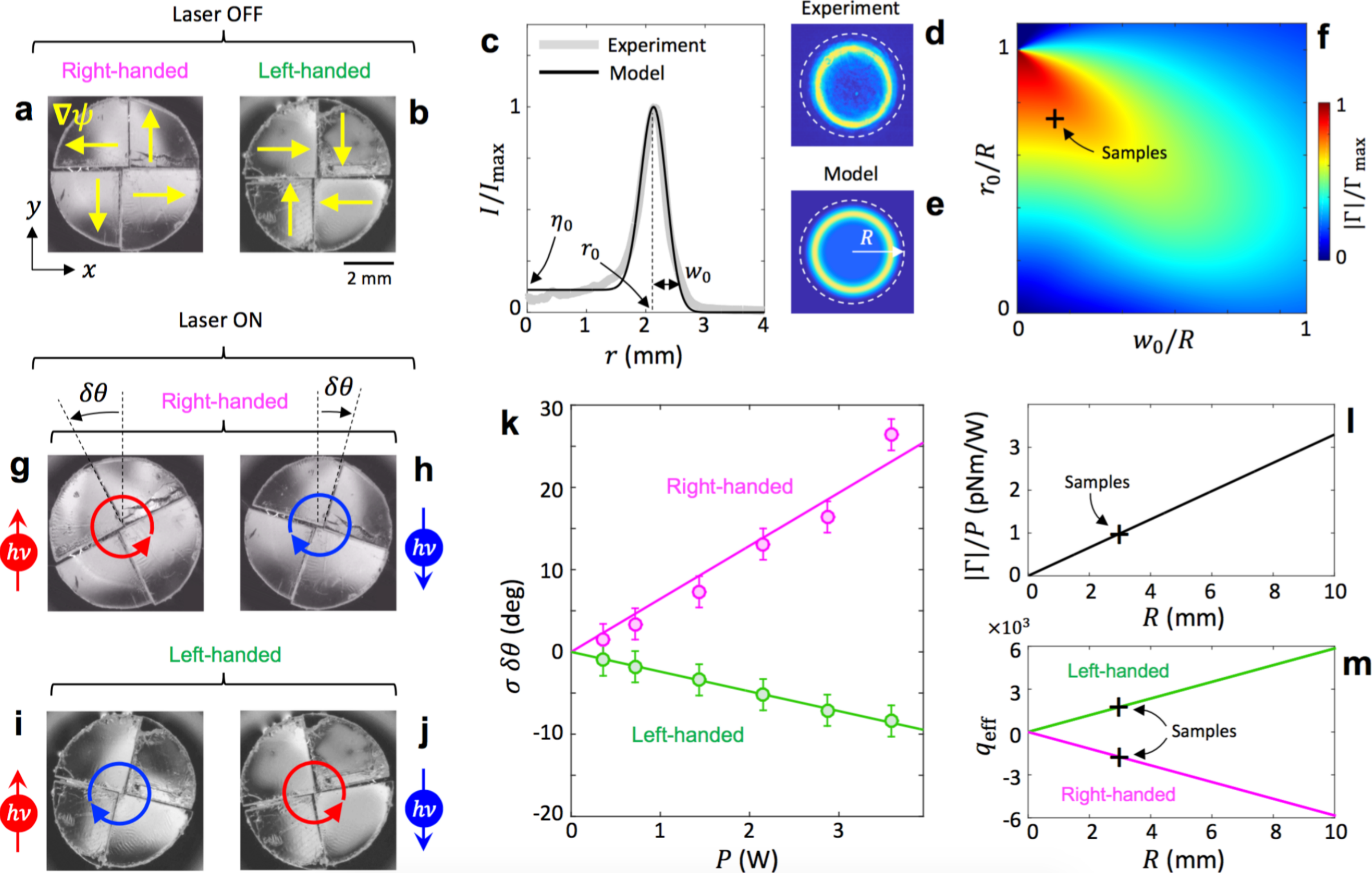}
\caption{
{\small {\bf Macroscopic direct observation of the angular analog of lateral optical forces.}
(${\bf a}$,${\bf b}$) Disk-shaped capillarity-bound four-quadrant spin-orbit rotors with $R \sim 3~$mm radius made of hand-cut large area cycloidal diffractive waveplate (see Fig.~2) placed at the air-water interface of a $2~{\rm cm}$ diameter circular plexiglass pool with 5~mm depth. Two samples with opposite orientational gradients ${\boldsymbol \nabla} \psi$ are prepared to ensure right-handed (${\bf a}$) and left-handed effects  (${\bf b}$). (${\bf c}$) Azimuth-averaged ring-shaped experimental radial intensity profile in the plane of the sample and its adjustment using the ansatz function given in the text, which gives $\eta_0=0.12$, $r_0 = 0.73R$ and $w=0.15R$ values for the three adjustable parameters. The experimental and calculated transverse intensity profiles are shown in panels (${\bf d}$) and (${\bf e}$), where the dashed circle refer to the outer part of the disk structure. (${\bf f}$) Calculated normalized optical torque magnitude on an ideal four-quadrant disk irradiated by circularly polarized incident light using as a function of the reduced parameters $r_0/R$ and $w/R$ at fixed power and imposing $\eta_0=0.12$, where $\Gamma_{\rm max}$ refers to the maximum of $|\Gamma|$ . The cross marker refers to the sample characteristics. (${\bf g}$)$-$(${\bf j}$) Typical steady-state angular deviation $\delta\theta$ associated with right-handed (${\bf g}$,${\bf h}$) and left-handed (${\bf i}$,${\bf j}$) optical torque experimental for incident spin up and spin down photons and total incident optical power $P \sim 3.5$~W. (${\bf k}$) Power dependence of $\sigma \delta \theta$ whose sign allows straightforward identification of the right-handed and left-handed nature of the applied optical torque. (${\bf l}$) Calculated torque per unit power as a function of the radius of an ideal four-quadrant disk irradiated by circularly polarized incident ring-shaped beam in the sample plane using the actual beam parameters. (${\bf m}$) Calculated structural topological charge $q_{\rm eff}$ of an equivalent effective q-plate with radius $R$, see text for details.}
}
\end{figure}

Qualitative observations for spin up and spin down incident photons for both right-handed and left-handed situations at fixed incident power are displayed in Figs.~5(g)$-$(j), that show static angular deviations by an angle $\delta \theta$. We note that the absence of light-induced spinning indicates the existence of capillary torsion spring effect that is likely due to uneven spatial distribution of the capillary effect associated with non-ideal circular shapes of the home-made spin-orbit optical structures. Quantitative power dependence of the effect is shown in Fig.~5(k), where right-handed and left-handed nature of the driving optical torque is inferred by looking at the sign of the product $\sigma \delta\theta$ where $\sigma=\pm 1$ is the incident helicity ($\sigma=+1$ for spin up and $\sigma=-1$ for spin down), since ${\bf\Gamma} \cdot {\bf S}$ and $\sigma \delta\theta$ have the same sign. In addition, we emphasize the large value of the optical torque per unit power estimated from $| \Gamma|/P =\frac{2R\sin \theta}{\pi c}\int_0^R I(r) r^2 dr \big /\int_0^\infty I(r) r dr$, that is of the order of $1~$pNm/W for our samples, as shown in Fig.~5(l). Finally, it is instructive to mention that present macroscopic samples behave as effective so-called q-plates \cite{marrucci_prl_06} with structural topological charges $|q_{\rm eff}| \sim 1700$, see Fig.~5(m). This is deduced by balancing above torque expression with that of the spin-orbit torque exerted on a q-plate irradiated by circularly polarized beam of power $P$, $2\sigma (1-q_{\rm eff})\frac{P}{\omega}$\cite{hakobyan_natphot_2014}, which gives $q_{\rm eff} = 1\mp \frac{|\Gamma| \omega}{2P}$ for the right-handed and left-handed cases, respectively.

These results unveil for the first time that direct experimental observations of unusual optomechanical manifestations are restricted neither to tricky instrumental approaches nor to weak transverse effects vanishing in the single-dipole limit. This points out that structured anisotropic matter is a wonderful playground for exploring new facets of light-matter interactions, which do not necessarily require the use of coherent light sources. Remarkably, the reported macroscopic translational and rotational motions underpin the potential of optomechanics to develop novel technique to unravel the mechanical properties of complex soft and possibly biological interfaces, such as cell membranes that are known to be anisotropic and inhomogeneous viscoelastic objects.
\vspace{0mm}

\newpage
\noindent
{\bf Acknowledgements}

\noindent
This study received financial support from CONACYT Mexico.\\

\noindent
{\bf Author contributions}

\noindent
H.M. realized the experimental set-up, conducted the experiments and analyzed the data, E.B. conceived the experiment, analyzed the data and supervised the project, E.B. wrote the paper with some input from H.M.\\





\end{document}